# Augmented Reality Applied to LEGO® Construction

AR-based Building Instructions with High Accuracy & Precision and Realistic Object-Hand Occlusions

Wei Yan
Department of Architecture
Texas A&M University
College Station, Texas, USA
wyan@tamu.edu

**ABSTRACT**

BRICKxAR is a novel Augmented Reality (AR) instruction method for construction toys such as LEGO®. With BRICKxAR, physical LEGO construction is guided by virtual bricks. Compared with the state-of-the-art, accuracy of the virtual - physical model alignment is significantly improved through a new design of marker-based registration, which can achieve an average error less than 1mm throughout the model. Realistic object occlusion is accomplished to reveal the true spatial relationship between physical and virtual bricks. LEGO players' hand detection and occlusion are realized to visualize the correct spatial relationship between real hands and virtual bricks, and allow virtual bricks to be "grasped" by real hands. The integration of these features makes AR instructions possible for small-parts assembly, validated through a working AR prototype for constructing LEGO Arc de Triomphe, quantitative measures of the accuracies of registration and occlusions, and heuristic evaluation of AR instruction features.

## 1 INTRODUCTION

Enabling spatial play, construction toys serve an important role as metaphors for scientific principles, spatial skills and math aptitude [14,24,29,42]. The improvement of the performance of construction toy players in terms of building time and number of building errors contributes to the players' overall learning [11]. The performance heavily depends on the instructions that guide the construction process. There are two primary tasks in designing the instructions: (1) Planning a sequence of assembly operations for users to understand and follow easily, and (2) Presenting the assembly operations clearly in a series of diagrams [3]. This project has developed an Augmented Reality (AR)-based instruction method: BRICKxAR. With BRICKxAR, physical toy construction (LEGO as an example) is guided by virtual bricks in the right place at the right time, step by step. Information augmentation is created for specific bricks about architecture and construction knowledge. Physical and virtual object occlusion is implemented to enable a natural appearance of virtual bricks on the physical model. Players' hand detection and occlusion are accomplished to allow a realistic immersive AR experience, in which virtual bricks can be "grasped" by the real hand, revealing correct spatial relationship of objects.

In BRICKxAR, the physical model can be moved and rotated freely on a desk surface, and the AR device camera can move in 6-Degrees of Freedom (DoF); in the meantime, high accuracy of AR registration - the virtual model's alignment with the physical model - is achieved through Computer Vision-powered marker-based registration using camera and motion sensors (including gyroscope, accelerometer, and magnetometer). The average error of the registration is less than 1mm throughout the entire model when a major part or full of the marker is within the AR camera's field of view. Compared with the state-of-the-art of AR instructions, BRICKxAR significantly improved accuracy of model registration. To validate the project contributions, LEGO Architecture 21036 Arc de Triomphe in Paris is built completely for its 386 steps with a working prototype of BRICKxAR in experiments for quantitative measures of the accuracies of registration and occlusions, and for heuristic evaluation of AR instruction features by comparing BRICKxAR with design principles of assembly instructions and AR design guidelines suggested by literature.

## 2 BACKGROUND AND LITERATURE REVIEW

Major assembly and construction projects are increasingly complex [45]. In recent years, LEGO sets also become more complex and labor intensive [27]. Some sets have thousands of pieces, e.g. LEGO 75192 Star Wars Millennium Falcon has 7,541 pieces and an instruction booklet of 496 pages. In the meantime, Augmented Reality (AR) can superimpose digital images on the real world view of users, having the potentials to benefit manufacturing, building construction, and part assembly significantly. AR has been studied as education and research tools, e.g., in the interactive AR assembly guidance and learning of Tou-Kung (a sophisticated cantilevered bracket system in traditional East Asian architecture) [10], in tutoring machine tasks with AR-assisted spatial interactions [9], and in the research of spatial design and urban planning problems by projecting data visualization onto LEGO models [4]. The LEGO Group uses AR to mix touch-screen game components, such as audiovisual effects, with the physical models, in order to make LEGO play more engaging and fun [26]. However, one of the core LEGO building experiences – the construction process – may also be enriched from the transformational AR technology, which is the goal of BRICKxAR.

### 2.1 Benefits and Limitations of 3D-Model-Based AR Instructions

A recent study about a 2D projection-based AR-assistive system shows only small benefits in the training scenario, in which AR training does not reach the personal training in terms of speed and recall precision after 24 hours [8]. However, many studies have verified that 3D-model-based AR instructions can significantly save task completion time [20] and reduce the error rate [35,41]. Funk et al. proposed using Duplo (a larger version of LEGO with each dimension doubled) and an artificial industry task for evaluating AR instructions in terms of assembly time and errors, with paper-based instructions as baselines [16]. Westerfield et al. evaluated participants' performance for assembly of computer motherboards, and found significant improvement using AR with a feedback system [44]. They also pointed out the drawback of limited accuracy in AR tracking. Schwald and de Laval presented an AR prototype for assisting in equipment maintenance and concluded that the results were positive, but improvements were needed for the accuracy of 3D augmentation [36]. Tang et al. compared an AR-based instruction system with a printed manual, computer-assisted instructions on a monitor, and a Head-Mounted Display for Duplo assembly [40]. Their user evaluations support the proposition that AR systems improve assembly performance significantly. They found, however, the limitations of tracking and calibration techniques being the biggest obstacles. Similarly in [21], research has confirmed the benefits of using AR instructions over traditional digital or paper instructions, and found that even though HoloLens AR instructions are a preferable alternative to traditional, 2D assembly instructions and tablet-based AR, the position of the virtual objects in HoloLens AR did not align correctly with the assembly parts.

### 2.2 AR Registration

In 3D-model-based AR instructions, registration accuracy (on the mean localization) is required based on the actual tasks. To enable small parts assembly (such as LEGO) using AR instructions, improving model registration is needed. For example, many LEGO bricks are very small in size: 1 LEGO Unit is 1.6mm (the thickness of the plastic wall) and the stud's diameter is 3 units (= 4.8mm). Low accuracy will result in significant misalignment between virtual and physical models, and thus errors in construction. A comprehensive review of AR for assembly points out that accuracy and latency are the two critical issues [43]. In a wood structure construction project using AR, the virtual object overlaying on the physical component causes uncertainty around positioning; also, the virtual object is not stable but shaking [34]. Using GearVR and a marker-based approach for constructing a timber structure has errors of 10-50mm between virtual and physical objects [2]. A recent workshop presents a HoloLens AR platform that enables interactive holographic instructions with a prototypical project to design and construct a pavilion from bent mild steel tubes [23]. The digital design model and the digitized physical model differ by at most 46 mm, with an average of 20 mm across all parts, attributed to human errors in construction, physical model self-weight and contortion, and holographic drift from inside-out device tracking [23]. In another collaborative construction project, some construction workers wear the HoloLens to instruct others without the device to build the structure, but the localization has an accuracy problem [18]. One of the findings from a survey of AR in the manufacturing industry in the last decade is that the marker-based solutions are usually the preferred tracking technology due to ease of implementation and higher accuracy compared with marker-less solutions [7]. A recent marker-based registration for AR's application with a tangible user interface in building design is demonstrated in [39].

### 2.3 Object and Hand Occlusions

Object occlusion is critical for correct depth perception in AR to ensure realistic and immersive AR experiences, but existing occlusion methods suffer from various limitations, e.g. the assumption of a static scene or high computational complexity [12]. An example is the usage of geospatial data to construct the geometric model of buildings in an outdoor environment to simulate occlusion between virtual content and real buildings in AR. While the simulated occlusion is achieved, the outcome is not realistic [25].

Model-based object occlusion has been studied for various AR applications, but there are few studies of object occlusions in AR instructions. Occlusion-by-Contours was found to aid users by removing ambiguity about depth perception using contour rendering for virtual parts occluded by physical parts in assembly [28] . However, the resulting images of the work demonstrates a registration accuracy problem and that the contour lines designed for the occluded portion of the virtual parts are not realistic for revealing an accurate spatial relationship between the physical and virtual parts. Thus, accurate object occlusion is still challenging to achieve for revealing the correct spatial relationship between physical and virtual parts in AR instructions.

Similar to the need for object occlusion, hand/people occlusion research is found in prior AR literature [1]. A relatively recent study is on optimizing the consistency of object boundaries between RGB and depth data obtained by an RGB-D sensor. While the approach achieved accurate hand occlusion, the performance is near real-time on a tablet platform, i.e. 30FPS (Frame Per Second) for a screen resolution of 640x480, and roughly 15FPS for 720P [12]. Still, there is a lack of research on hand occlusion in AR instructions.

Compared with the current literature on existing AR applications and features reviewed in this section, BRICKxAR has achieved significant improvement in AR instructions for (1) high accuracy with a high frame rate in AR registration, (2) realistic physical-virtual object occlusion, and (3) real-time accurate visualization of hand occlusion / real hands grasping virtual objects. The integration of these improved features makes AR instructions for small-parts assembly such as LEGO construction possible. The research methodology consists of prototyping, implementation, and evaluation.

## 3 PROTOTYPING AND IMPLEMENTATION

The prototyping process in BRICKxAR research includes virtual model preparation, marker-based registration, step-by-step instructions, object and hand occlusions, and implementation of the prototype – an app on AR-enabled iOS device.

### 3.1 Virtual Model Preparation

A virtual model for AR needs to be prepared with CAD modeling tools. LEGO Arc de Triomphe is used as an example due to the availability of both the physical set and its virtual model, as well as the moderate complexity of the set. The virtual model can be rendered in shaded and wireframe modes in AR (Figure 1).

Construction steps need to be introduced to the builder in a logical order. Thus, virtual bricks are stored in the order using an indexed array. In general, a brick at the lower elevation will be assembled before that at the higher elevation. However, in some cases, there are bricks to be attached to an upper brick instead of a lower brick, when the upper one acts like a bridge and the new brick needs to be attached to its bottom.

### 3.2 Marker-based Registration

Images can be set up as AR markers for the alignment of virtual and physical models. To transform the virtual model to the right scale, location, and orientation for accurate alignment between the virtual and physical models, different design options of markers were tested using images inspired by AprilTags [33]. A most accurate and robust registration was achieved by combining two images into a larger, single marker, with black color pixels filling the empty areas (Figure 2 left). It is also very flexible in that even if part of the marker is covered (for example, the upper-left black color area is gradually covered by the LEGO set step by step), at each step, the image marker is detected and tracked with high accuracy (Figure 2 middle and right), evaluated in Section 4.

### 3.3 Step-by-Step Instructions

The virtual brick of each current step is rendered in the right place of the model to guide the player for physical construction (Figure 3). During the construction process, important architecture information is displayed on the user interface for specific, relevant steps. For example, at Step 137, the photo of the sculpture - Napoleon crowned by the goddess of victory and text explanation are shown and linked to the LEGO micro-figure that represents the sculpture (Figure 4).

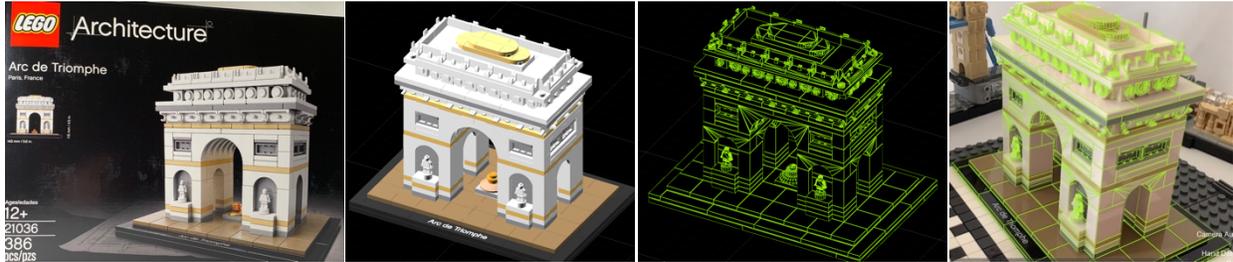

Figure 1. Left to right: 1) LEGO Arc de Triomphe, 2) and 3) shaded and wireframe virtual models, and 4) In BRICKxAR, the virtual model (green wireframe) aligns with the physical model.

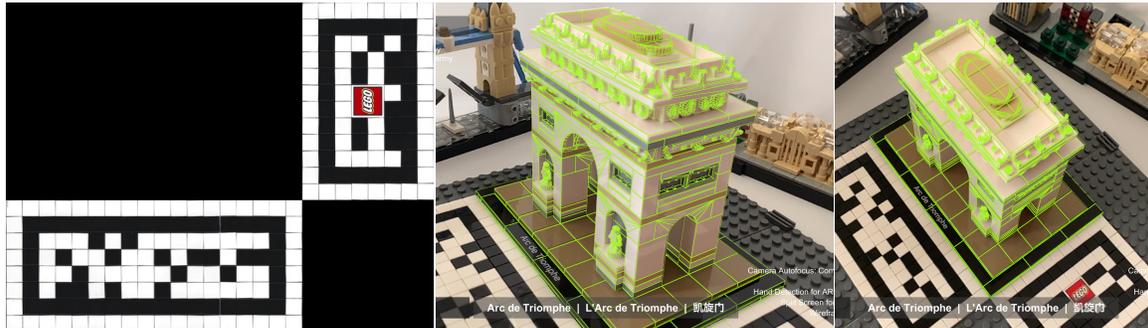

Figure 2. Left: a scanned image of the LEGO pieces-made plate serving as the marker for AR registration. Middle and right: high accuracy of BRICKxAR registration. The virtual model (green wireframe) aligns accurately with the physical model.

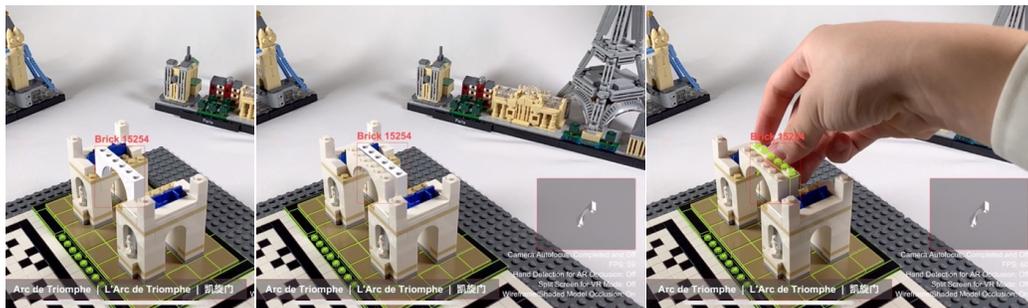

Figure 3. BRICKxAR: shaded or wireframe virtual bricks guiding physical LEGO construction. Left and middle: 1st and 2nd virtual Bricks 15254 (below the red labels), and right: player assembling the 2nd physical Brick 15254.

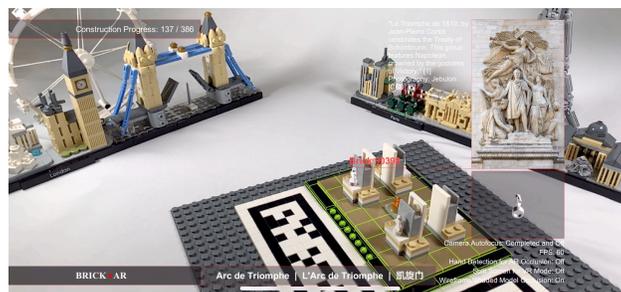

Figure 4. Step 137: photo and text information about the sculpture - Napoleon crowned by the goddess of victory - are shown on the upper-right, and linked to a LEGO micro-figure (Brick 90398).

## 3.4 Physical-Virtual Object Occlusion in Step-by-Step AR Instructions

The algorithm performing object occlusion in the step-by-step BRICKxAR instructions is shown below.

---

ALGORITHM 1: Step-by-Step Object Occlusion

---

start AR session

current_step = first_step

capture image marker with AR camera

determine a transformation for current_step virtual brick based on marker-based registration

render current_step virtual brick using Shader #1

guide user to assemble a corresponding current_step physical brick using current_step virtual brick

change the rendering of the current_step virtual brick using Shader #2

previous_step = current_step

current_step = next_step

while current_step ≤ final_step, do

    for each frame update, do

        capture image marker with AR camera

        determine a transformation for current_step virtual brick based on marker-based registration

        render current_step virtual brick using Shader #1 to appear being occluded by previous_step physical brick (and any previous physical bricks)

    end

    guide user to assemble a corresponding current_step physical brick using current_step virtual brick

    change the rendering of current_step virtual brick using Shader #2

    previous_step = current_step

    current_step = next_step

end

---

In the algorithm, Shader #1 is a shaded or partially transparent wireframe rendering function for depicting the transformation of the current virtual brick in the AR image to guide construction. Shader #1 allows the current step's virtual brick to be visible fully or partially depending on whether there are existing physical bricks in front of it or not. Shader #1 is made through a combination of shader properties, including:

- Rendering mode: fully opaque or transparent for mesh faces but opaque for edges
- Cull: back, determined by depth testing in the camera space

Shader #2 is a transparent and occlusive rendering function for showing the previous physical brick after it is inserted. It makes physical bricks appear to occlude a virtual brick behind them. Shader #2 is made through a combination of shader properties, including:

- Rendering mode: transparent (for both mesh faces and mesh edges)
- Cull: back, determined by depth testing in the camera space

## 3.5 Hand Occlusion and Real Hand Grasping Virtual Objects

Hand detection and occlusion are implemented to enable "grasping virtual objects with real hands", potentially enhancing realistic immersive AR experience of players. Hand occlusion in AR instructions becomes necessary in situations such as Step 350 (Figure 5), where the virtual brick appears to be in front of the fingers, demonstrating an incorrect spatial relationship between the virtual brick and the hand.

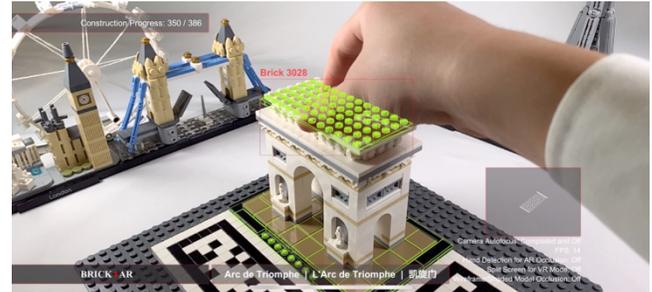

Figure 5. Step 350: virtual Brick 3028 appears in front of the hand.

Using Computer Vision, the hand area can be detected. Then virtual objects can be inserted into the scene to cover the hand area and rendered to occlude the virtual bricks, while being transparent to reveal the hand.

### 3.5.1 Hand Detection

Color segmentation is used for detecting hands by the skin colors. The target skin colors can be selected in real time by the player touching the hand area on the AR screen multiple times. The colors at the two most recent touched points are used as target hand colors, each with a predefined tolerance value. A grid of points on screen are compared with the target colors and labeled as hand points if their color values fall into any of the target color ranges. The grid density can be adjusted.

A "flood-fill" method is used to fill in the holes if they exist within the detected hand points. This will enable the entire hand area to be filled with Hand Occlusion Objects, in case the color detection result leaves any holes because of inconsistent colors on hands due to various reasons, e.g. lighting. In addition, very small blobs resulted from color segmentation are removed as they are likely to be other objects with similar colors as hands.

In Computer Vision, the YCbCr color space is suggested for effective and efficient performance of skin color segmentation [37]. BRICKxAR is able to access the Cb and Cr channels in the video frames captured by the AR device camera (iPhone's rear camera) in real time. Therefore the Cb and Cr channels are used in hand color segmentation based on the video frames.

### 3.5.2 Hand Occlusion

Hand Occlusion Objects (small 2D hexagons) are instantiated to cover all the detected hand points (within the point grid on screen) and the area around each hand point. The objects are transformed from the Screen Coordinate System to the World Coordinate System to be located in between the camera near-clipping plane and the virtual bricks. These hexagon objects are rendered in such a way that they are transparent in order to show the hands, while occluding the virtual bricks (Figure 6).

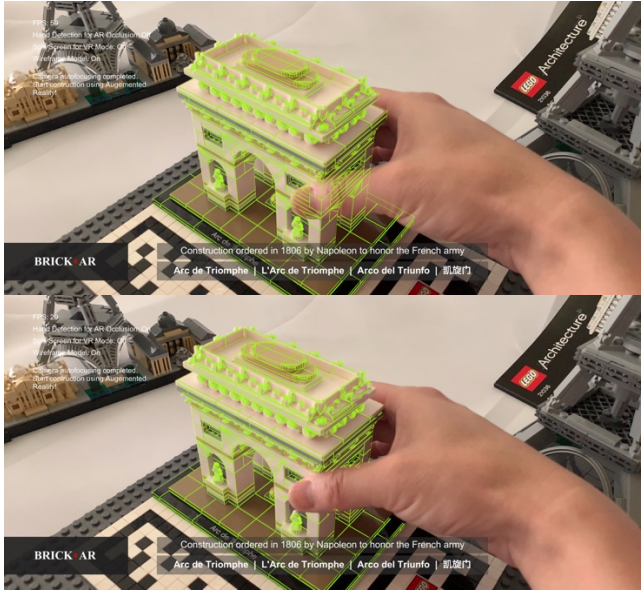

Figure 6. Top: without hand occlusion, the virtual bricks (wireframe or otherwise shaded) are not "grasped" by the hand, but appearing in front of the hand. Bottom: with hand occlusion, the virtual bricks are "grasped" by the real hand, just like the physical bricks.

In BRICKxAR, the process of hand occlusion is shown in ALGORITHM 2. The depth relation between the camera and the registered physical-virtual models is always known in the AR scene. When a hand is in front of the physical model, the hand will be detected and rendered in between the camera and the registered virtual model, to occlude the model. If the hand is behind the physical model, the hand will not be detected, therefore will not occlude the virtual model. This algorithm utilizes the accurate physical-virtual model registration. The interplay between physical and virtual models in BRICKxAR makes the problems and solutions of occlusion different from prior art of AR.

ALGORITHM 2: Real Hand – Virtual Object Occlusion

start AR session
for each frame update, do
    capture image marker with AR camera
    determine transformation of virtual bricks based on marker-based registration
    render virtual bricks using Shader #1 (see 3.4)
    detect hand area in the image
    generate virtual hexagon objects to fill the detected hand area
    render virtual hexagon objects with Shader #2 (see 3.4), wherein virtual bricks appear to be occluded by real hands
end

### 3.5.3 *Real Hand Grasping Virtual Bricks*

Compared with literature, BRICKxAR has achieved not only accurate occlusion in real time, but also the realistic visual effect of "grasping" virtual objects, which is enabled partially by hand occlusion and partially by the actual grasping of the virtual model's counterpart – the physical model (Figure 6 bottom).

### 3.6 Implementation

BRICKxAR's software architecture diagram is shown in Figure 7. For implementing the BRICKxAR prototype, AR hardware and software were reviewed and iPhone XS Max, Apple's ARKit, the Unity game development software, and UnityARKitPlugin were selected due to iPhone's advanced camera and motion sensors accessible from ARKit, Unity's interactive gaming and graphics capabilities, and UnityARKitPlugin enabling Unity games to interact with ARKit. C# was used for Unity programming and Objective-C for the iOS app development.

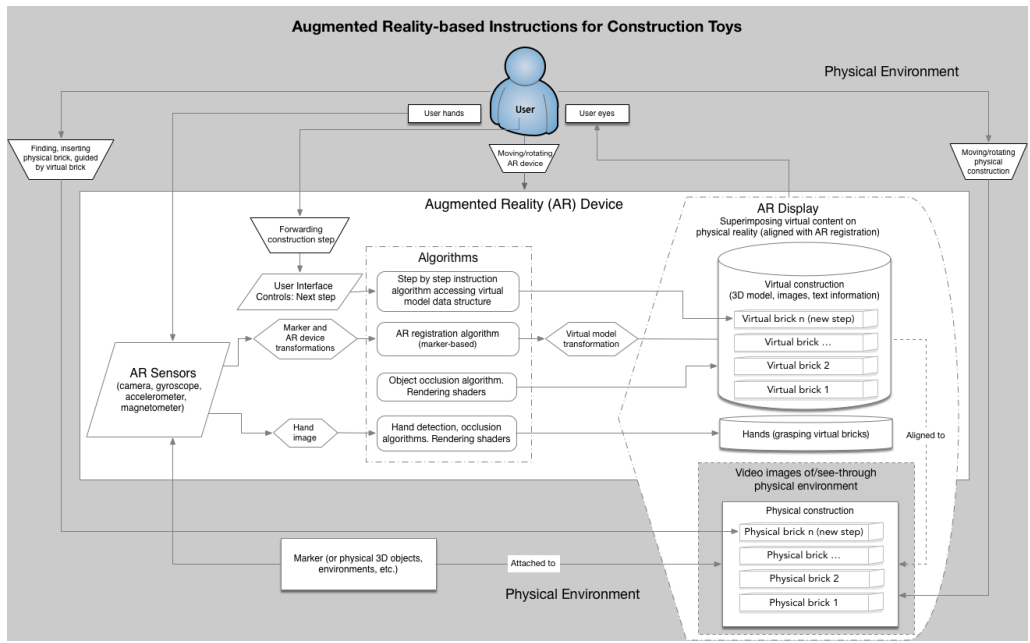

Figure 7. BRICKxAR's software architecture diagram

# 4 EVALUATION AND VALIDATION

3D-model-based AR has proven in literature for improving assembly performance significantly [20,35,40,41], but accuracy and latency are the critical issues [43], and the lack of concrete design guidelines of AR applications is a key barrier [5]. Based on these findings, the validation of the BRICKxAR research is designed as follows: an example LEGO set (Arc de Triomphe) has been built completely with BRICKxAR for all the 386 steps in experiments for (1) quantitative measures of the accuracy of model registration, compared with the state-of-the-art from literature, (2) quantitative measures and visual examination of the accuracies of object and hand occlusions, and (3) heuristic evaluation of BRICKxAR features for enhancing instruction design, compared with major design principles of assembly instructions and AR application design guidelines, both suggested by literature.

## 4.1 Improvement of Registration Accuracy

To evaluate the registration accuracy comprehensively, the LEGO Arc de Triomphe set has been built completely with BRICKxAR and the process was recorded using the AR camera (see video demo [46]). By examining the complete video of a total length 1h 34'51", at a resolution of 2336x1080 pixels (different from the AR screen resolution 2689x1242), at 29.92 FPS, the average error of the registration is found to be less than 1mm throughout the entire model, when at least the major area of the marker is within the AR camera's field of view. For example, Figure 4 shows the camera's entire field of view and the marker is partially out of the view, but the registration is still accurate, as seen with the alignment between the green wireframe virtual tiles and the physical tiles. Error is defined as the distance between the corresponding edges on the virtual and physical models, obtained through visually examining the video images and measurement on sample images. The error is affected by the following factors: (1) the fidelity of the original CAD model, (2) the attachment of physical bricks being tighter or looser with slight, free rotations made by a player, (3) ARKit's image tracking algorithm, and (4) BRICKxAR marker design and parameter settings, e.g. the marker's claimed size, which can be calibrated. The registration error propagates and increases from locations closer to the AR device camera and the marker to locations farther away. The marker image size also contributes to the error: the smaller size, the bigger error.

If the major area of the marker is covered (resulting in the loss of AR tracking), BRICKxAR will fix the virtual model in the latest tracked location. If at this time the physical model is moved or rotated, the registration error will increase from less than 1mm to misalignment in millimeters then centimeters, and to the worst case where the virtual and physical models become apart (the drifting effect). However, as soon as the major area of the marker is uncovered inside the camera's field of view, i.e. the most cases, highly accurate registration and tracking will resume immediately. In another experiment, a much larger LEGO set 10243 Parisian Restaurant containing 2,469 bricks verifies the high accuracy of registration again (Figure 8). The performance was shown in different indoor environmental lightings - different rooms with different natural and artificial lighting conditions during the development and demonstrations of the project. Compared with the state-of-the-art (Section 2), significantly improved accuracy of registration is achieved in BRICKxAR.

The large marker design with two patterned areas improves the marker-based registration accuracy, so that AR-based LEGO construction is made possible. In addition to the marker-based approach, 3D point cloud SLAM (Simultaneous Localization and Mapping [6,13]) was also tested in BRICKxAR, but the accuracy was not sufficient for the purpose of LEGO construction. However, if in the future a high-resolution true-depth camera or LiDAR can be used for understanding the physical LEGO model's 6-DoF poses, accuracy for markerless tracking can be further investigated.

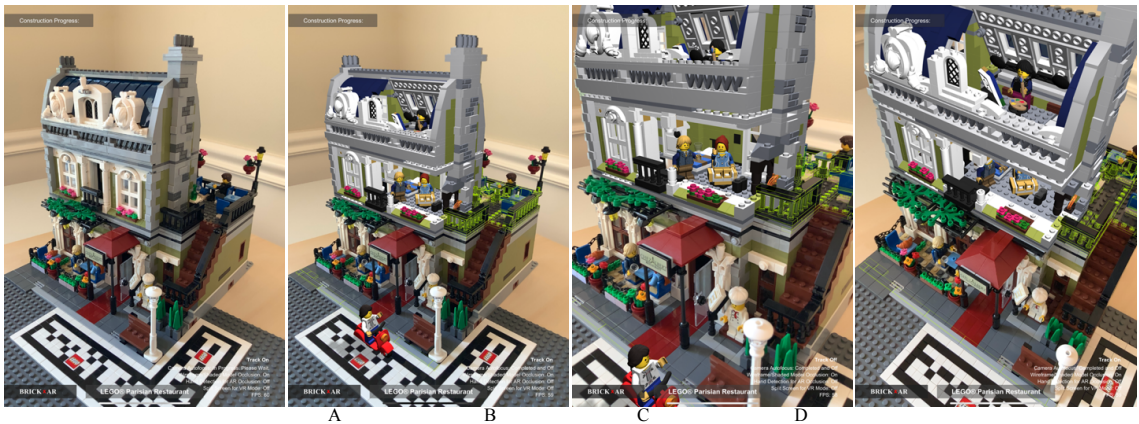

Figure 8. A: physical LEGO Parisian Restaurant model. B: virtual model of the 2$^{nd}$ and 3$^{rd}$ floors superimposed on the physical model. C and D: virtual cutaway views for revealing the hidden structure of the 2$^{nd}$ and 3$^{rd}$ floors in BRICKxAR.

## 4.2 Accuracy of Object and Hand Occlusions

**The accuracy of object occlusion** is measured by examining the entire 386 steps in the experiment of constructing LEGO Arc de Triomphe with BRICKxAR and watching the recorded video. During each step, when the AR camera moves around the LEGO, the physical-virtual brick occlusion works correctly, showing their realistic and accurate spatial relationship to enable instructions. For example, in Figure 9, virtual Brick 87079 appears realistically, similar to the real Brick 87079 that is closer to the camera, in terms of object occlusions. The virtual brick occludes the physical bricks behind it and is partially occluded by the physical bricks in front of it. The same realistic and accurate occlusions can be seen in Figure 3 for virtual Bricks 15254. 50 to 60 FPS at screen resolution 2689x1242 pixels is achieved when running the AR instructions with object occlusion.

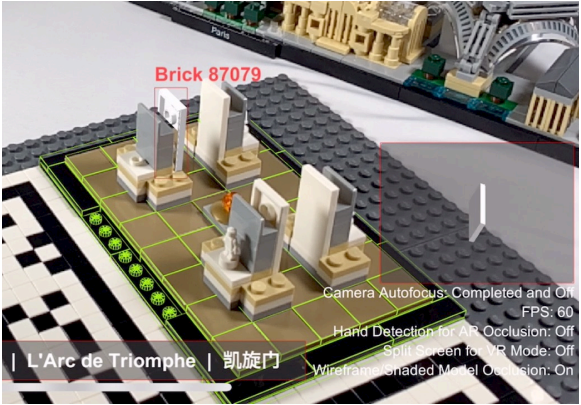

Figure 9. Step 136: virtual Brick 87079 (below the red label) appears realistically in terms of occlusions

**The accuracy of hand occlusion** is measured on sample images from the recorded AR video using Intersection over Union (IoU) – a widely-used method for evaluating image segmentation models:

$$IoU = \frac{Intersection\ Area}{Union\ Area}$$

where *Intersection Area* is the number of pixels in the intersection between the manually-selected hand area and the algorithm-detected hand area, and *Union Area* is the number of pixels in the union of the two areas. The manually-selected hand area is the ground truth. The algorithm-detected hand area can be automatically rendered in BRICKxAR as blue pixels for visualizing the hand detection results, as shown in Figure 10. The intersection and union of the two areas are shown in Figure 11. The IoU of hand occlusion equals to 88.3% on the sample images. After turning on hand detection and occlusion, the performance of the AR session is real-time, between 28-60 FPS on iPhone XS MAX's high resolution screen (Table 1), which can be compared with the near-real-time performance and lower resolutions in other studies, e.g. [12].

Table 1: AR experiment parameters

| Screen resolution (pixel) | Camera captured YCbCr image resolution (pixel) | | Recorded video resolution (pixel) |
|---|---|---|---|
| Screen Width x Height | Y-Channel Width x Height * | CbCr-Channels Width x Height | Video Width x Height |
| 2689 x 1242 | 1920 x 1440 | 960 x 720 | 2336 by 1080 pixels |

* YCbCr image height is higher than the screen height, as default by ARKit. Cropping extra pixels was made in the hand detection algorithm in BRICKxAR.

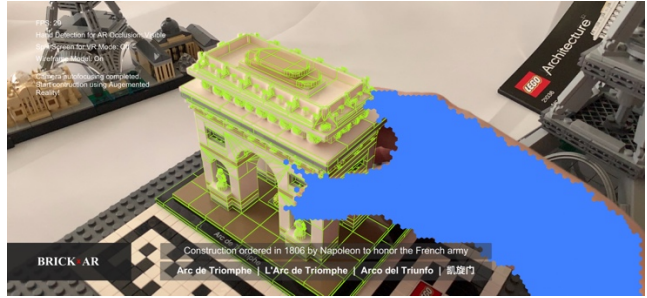

Figure 10. Hexagon objects are rendered in blue in the detected hand area to show the detection result. If the hexagons are rendered in a transparent yet occlusive material, the hand will appear occluding the virtual bricks, as in Figure 6 bottom.

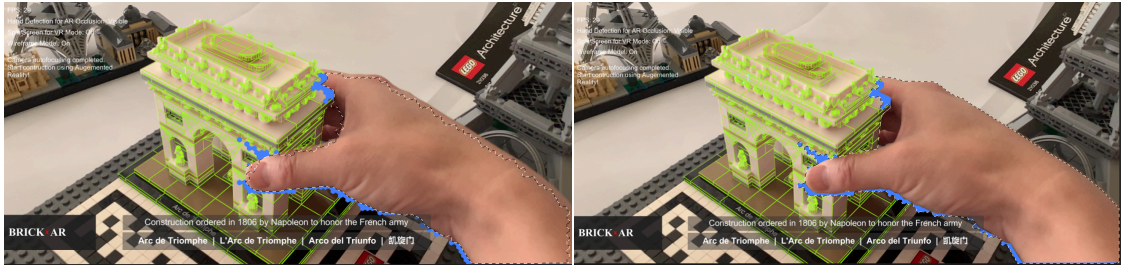

Figure 11. Both left and right: the manually-selected real hand-area (i.e. ground truth) is superimposed on the detected hand-area. Left: Intersection between the two areas (within the lasso) and right: Union of the two areas (within the lasso), for calculating Intersection over Union (IoU).

### 4.3 Heuristic Evaluation

Heuristic evaluation – comparison with rules of thumb is a Human-Computer Interaction research method [15,32]. BRICKxAR is compared with the major instruction design guidelines from literature and all the nine AR design guidelines proposed in recent literature [15]. The findings about BRICKxAR advancing the AR instruction methodology are explained below.

1. "*Current parts visibility (CPV). Each part in the current subset should be visible with respect to the other parts in the subset*" [19]. In BRICKxAR, the current part is the virtual brick. If the virtual brick is occluded by other parts (physical bricks), it can be made visible by the player rotating the physical model or moving the AR camera. The current part always has a correct spatial relationship with respect to other parts.

2. "*Previous parts visibility (PPV). Some portion of the parts attached in earlier steps should remain visible for context*" [19]. In BRICKxAR, parts in earlier steps are the physical bricks and visible naturally.

3. "*Future parts visibility (FPV). We want to ensure that parts added in an earlier assembly step do not occlude parts added in a later assembly step*" [19]. In BRICKxAR, the player can rotate the model or move the AR camera to reveal a future part (virtual brick) any time.

4. Instructions should support and build on the players' existing schemas and mental models to minimize extraneous cognitive load [30]. The choice of visualizations and metaphors should match the mental models of users based on their physical environment and task [15]. In BRICKxAR, the realistic shaded rendering, occlusions, and the aligned perspective views between the physical and virtual models match the player's level of graphic literacy and mental models naturally, which further support the form of virtual objects based on existing metaphors of users for communicating affordances and capabilities [15].

5. Alignment of physical and virtual worlds [15] is achieved with BRICKxAR's model registration.

6. In an instruction booklet, limited angles of view (mostly isometric) may obscure parts [31]. In BRICKxAR, the continuous changes of unrestricted viewing-angles and the perspective views are consistent with the player's natural graphics literacy, eliminating the obscuring problem.

7. In a previous study, when a digital model on a tablet was used as instructions, the rotational view was used often by players, much more than zoom and time-lapse views [11]. In BRICKxAR, rotation and zoom for the virtual model can be done by physically manipulating the LEGO model or the AR device, minimizing distraction and overload [15].

8. Model scale 1:1 is shown from time to time in the instruction booklets to distinguish bricks of similar shapes but different sizes, however, the graphic syntax of model scale could be confusing [29]. In addition, players had difficulty selecting some components correctly when the instruction colors did not accurately reflect the true colors of components [29]. For example, in some LEGO instruction booklets, "*orange parts appeared yellow, and black was depicted as a dark grey, presumably to allow edges and features to be shown in black*" [29]. In AR, to fit with user's perceptual abilities, designers should consider size, color, motion, distance, and resolution [15]. In BRICKxAR the virtual model's scale or size is automatically matched to the physical model due to correct model registration. The screen resolution used is high (Table 1). The completed bricks are the physical bricks with their natural colors, and the virtual brick can be rendered photo-realistically with little color mismatch.

9. For instruction booklets, a guideline is to depict the assembly on a work surface or relative to the player's position [31]. Similarly, an AR design guideline is to adapt user position and motion [15]. In BRICKxAR, this is achieved automatically by the player positioning the LEGO set on the desk surface, and then the virtual brick instructions appear relative to the player's position with a correct spatial relationship.

10. Minimal "look times" or "look duration" of gazing at the instructions in between gazing at the assembly are important measures for the success of instruction booklets [30]. In BRICKxAR, the player always looks at the virtual and physical bricks at the same time, thus the minimal "look times" and "look duration" can be achieved straightforwardly. In addition, all physical motion required should be easy in AR [15], while BRICKxAR doesn't require different physical motion from the conventional LEGO assembly.

11. With instruction booklets, users must compare two consecutive Structural Diagrams to infer which parts are to be attached [3,19], while in BRICKxAR, all the diagrams are consolidated into one single LEGO physical model under-construction and a virtual brick to be attached.

12. Line drawings of cutaway views or cross-sectional views can be used to show normally hidden or hard to see parts [38]. BRICKxAR can superimpose a virtual cutaway view on a physical model to help recall the hidden structure of the completed part of the assembly (Figure 8).

13. Accessibility of off-screen objects is suggested as an AR design guideline [15]. For a specific step of construction using BRICKxAR, the guiding virtual brick can be seen within the screen and is sufficient for the task without recalling other virtual bricks. A corresponding physical brick will need to be found off screen from the box of parts. The virtual brick on the assembly and its copy with animation (rotation) in the sub-window (Figure 9 right) are intended for helping find the physical brick off screen. Virtual bricks occluded completely or off-screen (as in Figure 8 C and D) can be visible by physically rotating the LEGO assembly or the AR device.

14. AR experiences should be designed to accommodate the capabilities and limitations of the hardware platform [15]. The BRICKxAR prototype takes the advantages of the AR-enabled iOS platform with its powerful sensors for SLAM and high resolution screen for display. The hardware limitation of being tablet-based, compared with advanced Head-Mounted Display systems [21], is addressed by the integration of improved features in BRICKxAR.

The evaluation also found the following unmet guidelines, which can be addressed with future development.

1. Skipping repeated instructions is suggested for the repeated actions when building sub-models in current instruction booklets [3]. This requires a "reference" to the repeated instructions for multiple sub-models. Currently, BRICKxAR provides the repeated instructions step by step, thus the hierarchical structure of the model is "flattened". If the instruction of the structural hierarchy is important for learning the modeling method, additional visualization of sub-models need to be added.

2. While instructions should build on the players' mental models to minimize extraneous cognitive load, they should also allow for some cognitive drag [30]. Compared with the current instruction booklets where 2D isometric CAD drawings are often used for guiding assembly, BRICKxAR eliminates the training opportunities for players to learn and understand 2D isometric drawings. However, serious games with different levels of challenges may potentially be built into BRICKxAR, and engaging spatial and STEM trainings are possible in the future.

## 5 CONCLUSIONS AND FUTURE WORK

The major **contributions** of the BRICKxAR project include the following:

(1) Compared with the state-of-the-art, accuracy of the virtual - physical model alignment is significantly improved through a unique design of marker-based registration, which can achieve an average error less than 1mm throughout the model.
(2) Realistic object occlusion is accomplished to reveal the true spatial relation between physical and virtual bricks.
(3) Hand detection and occlusion are realized to visualize the correct spatial relation between real hands and virtual bricks, and allow virtual bricks to be "grasped" by the real hands in AR.
(4) The integration of the above features makes AR instructions possible for small-parts assembly tasks, validated through the working prototype for constructing LEGO Arc de Triomphe.

**Future work** will be built on the following ideas. The marker-based AR in this project can be applied to many other LEGO sets' construction or assembly with small parts, but it may not be applicable to all block toys and various types of assembly or construction. For example, for LEGO Technic, the player needs to grasp parts, and translate / rotate them in 6-DoF for connecting the parts in hand. Investigating more comprehensive SLAM methods for more flexible yet accurate AR registration will continue to be necessary. BRICKxAR has not been tested with AR glasses. However, the design and techniques demonstrated by BRICKxAR can be applied to other AR devices through future development. Assembly part tracking is needed for error detection. An example tracking-enabled instruction system is DuploTrack [17]. However, LEGO bricks used in BRICKxAR are much smaller than those in DuploTrack: in the Arc De Triomphe model there are many 1x1 (stud) LEGO bricks (8mm x 8mm x 4.8mm), while DuploTrack used 2 x 4 (stud) bricks (32mm x 64mm x 24mm), therefore they are at different magnitudes. LEGO brick or small part tracking and assembly error detection present a very different and much more challenging problem for future research. The research demonstrated the potential application of hand occlusion in AR instruction process. The underlaying technology of hand detection can be switched to other technologies, e.g. Deep Learning in the future. AR has great potentials for enhancing spatial and STEM learning [22]. Future development of BRICKxAR and user studies may further reveal its potential applications in education.